\documentclass[twocolumn,prl,reprint,notitlepage,superscriptaddress]{revtex4-1}
\usepackage[latin9]{inputenc}
\usepackage{amsmath}
\usepackage{verbatim}
\usepackage{graphicx}
\usepackage{esint}
\usepackage{multirow}
\usepackage{braket}



\begin{document}

\title{Observation of density-dependent gauge fields in a Bose-Einstein condensate based on micromotion control in a shaken two-dimensional lattice}

\author{Logan W. Clark}

\affiliation{James Franck Institute, University of Chicago, Chicago, Illinois
	60637, USA}

\affiliation{Enrico Fermi Institute, University of Chicago, Chicago, Illinois
	60637, USA}

\author{Brandon M. Anderson}

\affiliation{James Franck Institute, University of Chicago, Chicago, Illinois
	60637, USA}

\author{Lei Feng}

\affiliation{James Franck Institute, University of Chicago, Chicago, Illinois
	60637, USA}

\affiliation{Enrico Fermi Institute, University of Chicago, Chicago, Illinois
	60637, USA}

\author{Anita Gaj}

\affiliation{James Franck Institute, University of Chicago, Chicago, Illinois
	60637, USA}

\affiliation{Enrico Fermi Institute, University of Chicago, Chicago, Illinois
	60637, USA}

\author{K. Levin}

\affiliation{James Franck Institute, University of Chicago, Chicago, Illinois
	60637, USA}

\author{Cheng Chin}

\affiliation{James Franck Institute, University of Chicago, Chicago, Illinois
	60637, USA}

\affiliation{Enrico Fermi Institute, University of Chicago, Chicago, Illinois
	60637, USA}

\begin{abstract}
	We demonstrate a density-dependent gauge field, induced by atomic interactions, for quantum gases. The gauge field results from the synchronous coupling between the interactions and micromotion of the atoms in a modulated two-dimensional optical lattice. As a first step, we show that a coherent shaking of the lattice in two directions can couple the momentum and interactions of atoms and break the four-fold symmetry of the lattice. We then create a full interaction-induced gauge field by modulating the interaction strength in synchrony with the lattice shaking. 
	When a condensate is loaded into this shaken lattice, the gauge field acts to preferentially prepare the system in different quasimomentum ground states depending on the modulation phase. 
	We envision that these interaction-induced fields, created by fine control of micromotion, will provide a stepping stone to model new quantum phenomena within and beyond condensed matter physics. 
\end{abstract}

\maketitle

Synthesizing gauge fields for cold atoms opens the door to investigate
novel quantum phenomena associated with charged particles in an electromagnetic
field \cite{Dalibard2011,Goldman2014}; examples include quantum Hall effects, topological matter and anyonic excitations. Many experimental approaches have been developed in the past years to introduce gauge fields, including rapidly rotating gases \cite{Madison2000,Abo-Shaeer2001,Schweikhard2004}, Raman transitions \cite{Lin2009,Lin2011}, laser-assisted tunneling \cite{Hirokazu2013,Aidelsburger2013}, and lattice shaking \cite{Struck2013, Parker2013}.

As charged particles in motion also generate electromagnetic fields, a complete simulation of the particle-field system should include the feedback of the matter to the gauge field \cite{Wiese2013}. Such a dynamical gauge field would enable simulation of important models in condensed matter \cite{spinliquids,Baskaran1988,Levin2005} and in high energy physics, as in Yang-Mills theories \cite{aitchison1989}. 
Many mechanisms have been proposed for introducing dynamical gauge fields in quantum gases
\cite{Banerjee2012,Cirac2010,Kapit2011,Zohar2011,Zohar2013,Tagliacozzo2013}, opening exciting directions for cold atom research.

On the way to dynamical fields, there is a great deal of interest in generating density-dependent (equivalently, interaction-induced) gauge fields in which the effective field depends on the arrangement of atoms \cite{Goldman2014}.
For example, such a field can be used to study new phase transitions \cite{Keilmann2011, Greschner2014}
and one-dimensional particles with anyonic statistics \cite{Keilmann2011,Greschner2015,Cardarelli2016,Strater2016}.
Proposals have suggested generating density-dependent gauge fields using light-matter interactions \cite{Edmonds2013,Edmonds2013b}, lattice modulation \cite{Keilmann2011,Greschner2015,Cardarelli2016,Strater2016}, or interaction strength modulation \cite{Greschner2014}.
Experimental realization, however, remains elusive.

Lattice shaking has recently emerged as a promising experimental tool for generating gauge potentials in cold atom systems \cite{Eckardt2017}, enabling exciting developments including topological bands \cite{Jotzu2014, Flaschner1091,Tarnowski2017}. In our recent work, lattice modulation at a frequency near-detuned to an inter-band transition induces a quantum phase transition in Bose-Einstein condensates, resulting in domain formation \cite{Parker2013}, roton excitations \cite{Ha2015}, and critical dynamics that are both universal \cite{Clark2016} and coherent \cite{Feng2017}. In this lattice shaking scheme, the superfluid remains long lived and the atomic interactions play an important role to establish the ordering of superfluid domains.

\begin{figure}
	\includegraphics[width=\columnwidth]
	{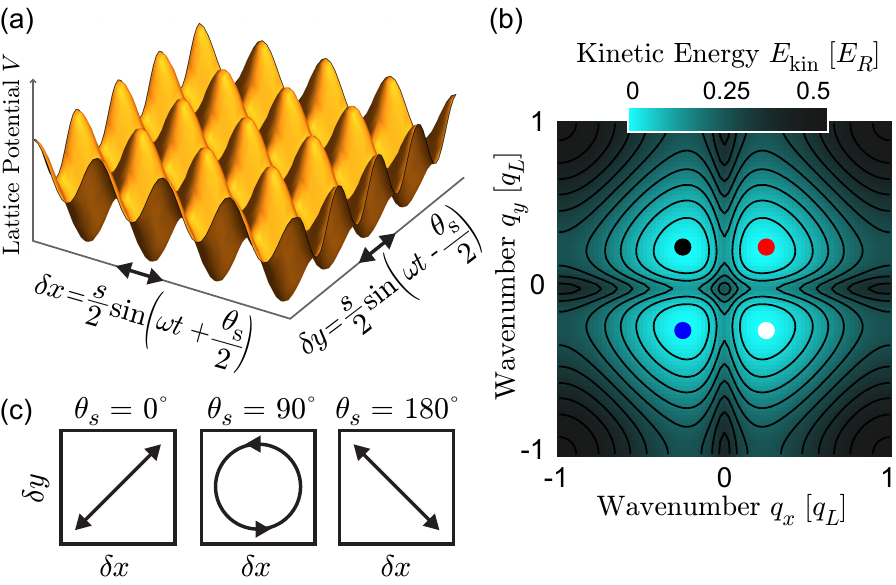}
	\caption{\label{fig:KEandMicro} Atoms in a two-dimensional shaken lattice.
		(a) A 2D, square lattice (orange surface) is shaken by inducing periodic displacements $\delta{x}$ and $\delta{y}$ along
		the $x-$ and $y-$axes respectively (arrows) with equal amplitude $s$ at frequency $\omega\equiv2\pi/\tau$, shaking period $\tau$ and relative phase $\theta_s$. (b) Shaking above the critical amplitude $s>s_{c}$ results in a single particle dispersion with four degenerate
		minima in the ground band at $\mathbf{q}=\left(+q^{*},+q^{*}\right)$,
		$\left(-q^{*},+q^{*}\right)$, $\left(-q^{*},-q^{*}\right)$ and $\left(+q^{*},-q^{*}\right)$,
		denoted respectively by red, black, blue, and white dots. 
		(c) The shaking phase $\theta_{s}$ controls the polarization of the lattice displacement. The polarization does not affect the single particle dispersion shown in (b).}
\end{figure}

In this paper, we experimentally demonstrate an interaction-induced synthetic gauge potential in a Bose-Einstein condensate. The gauge potential $\mathbf{A}(\rho)$ appears as the substitution, 
\begin{equation}\label{EqIIGP}
\mathbf{q} \rightarrow \mathbf{q}-\mathbf{A}(\rho)/\hbar
\end{equation}
in the Hamiltonian, linking its dependence on the momentum, represented by the wavevector $\mathbf{q}=(q_x,q_y)$, with  $\rho$, the density coarse-grained over one unit cell. 
Equivalently, one can view the interaction-induced field in a tight-binding model as an imaginary part of the tunneling  which depends on the occupation number operators $\hat{N}_k$ and $\hat{N}_{k+1}$ of the tunnel coupled sites,
\begin{equation}\label{eqn:PeierlsPhase}
J \rightarrow J + iJ'\left(\hat{N}_k+\hat{N}_{k+1}\right),
\end{equation}
where $J$ is the tunneling energy without the field and $J'$ encodes the strength of the density-dependent field \cite{supplement}.

To create this density-dependent gauge field we exploit the micromotion of atoms in a shaken 2D square optical lattice in combination with periodically modulated interaction strength.
For atoms condensed in a two-dimensional momentum state $\mathbf{q}$, this combination yields a mean-field energy shift,

\begin{equation}\label{Eq1}
\mathcal{E}_{\mathbf{q}}=  \eta_{\mathbf{q}} \rho g_0 ,
\end{equation}

\noindent where $g_0=\overline{g(t)}$ is the period-average of the interaction strength $g(t)=4\pi \hbar^2 a(t)/m$, $a(t)$ is the scattering length, $m$ is the atomic mass, and $2\pi\hbar$ is Planck's constant. The dimensionless interaction factor $\eta_{\mathbf{q}}$ accounts for the coupling between the micromotion and atomic interactions, as detailed below. A gauge potential in the form of Eq.~(\ref{EqIIGP}) requires $\eta_{\mathbf{q}}$ to be linear in $\mathbf{q}$.

We perform the experiment in two stages. In the first stage we show the effect of micromotion on interactions by tuning the relative phase $\theta_s$ between the lattice shaking in the $x-$ and $y-$directions while keeping the scattering length stationary. The micromotion raises the time-averaged interaction energy along the direction of shaking and can break the four-fold symmetry of the dispersion.
In the second stage we generate a density-dependent gauge field by modulating the scattering length with a phase $\theta_g$ relative to the lattice shaking. This scheme creates a gauge field with $\mathbf{A}\sim \mathbf{e}_{\Theta}\rho g_0 $, where $\mathbf{e}_{\Theta}$ is a unit vector in the direction $\Theta \equiv \theta_g -\theta_s/2$.
In both stages we test for the predicted effects via their influence on the phase transition in the shaken lattice.

\begin{figure}
	\includegraphics[width=\columnwidth]{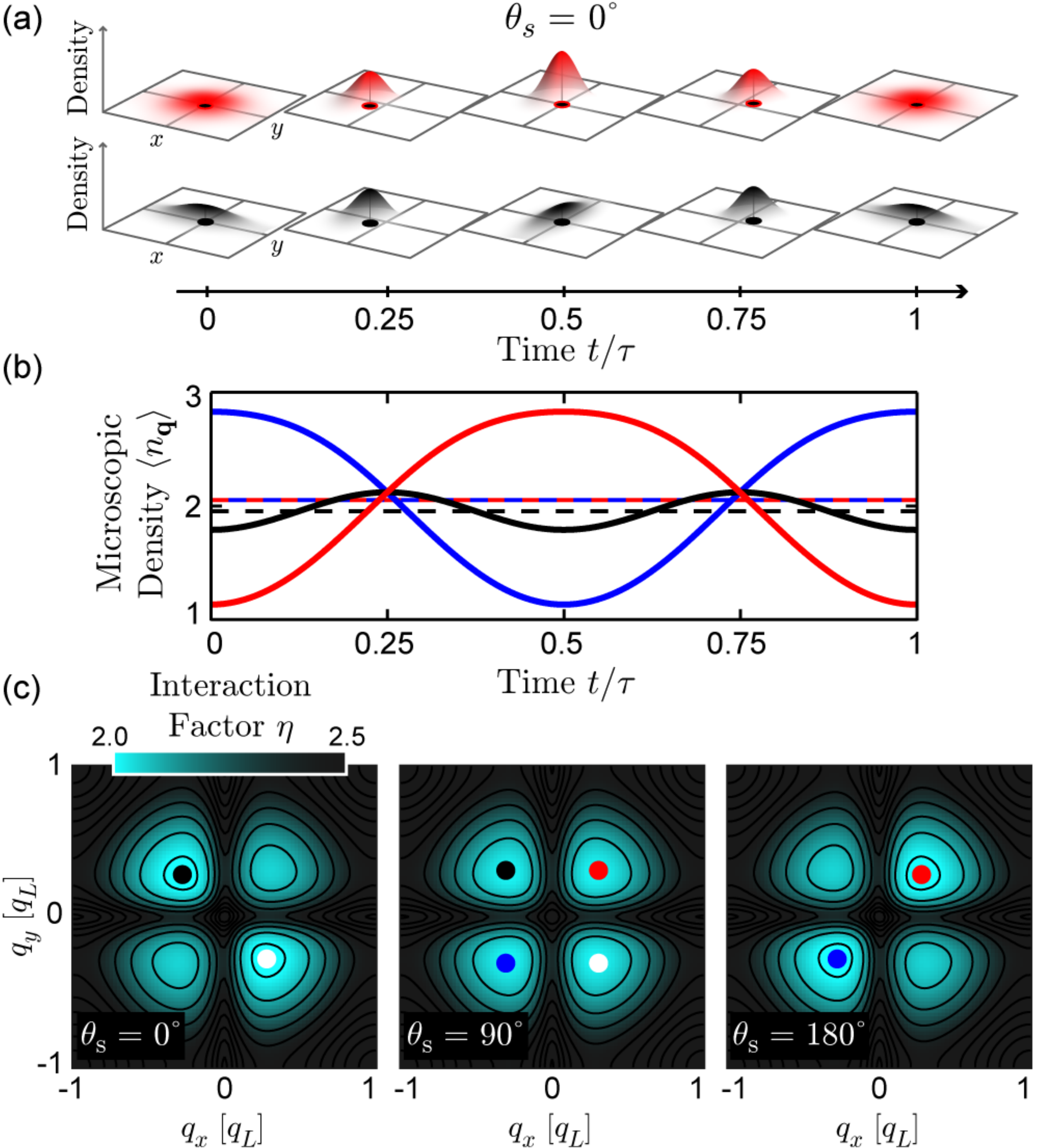}
	\caption{\label{fig:interactionsVsPolarization} Interaction-momentum coupling due to micromotion. (a) Examples of micromotion for linear shaking ($\theta_s=0^\circ$). Snapshots of the density $\left|\psi_{\mathbf{q}}\left(x,y,t\right)\right|^{2}$ within a single 2D lattice site are shown for two states, $\left(+q^{*},+q^{*}\right)$ (red) and $\left(-q^{*},+q^{*}\right)$ (black), within a shaking period $\tau$. (b) As a result of the micromotion, the mean microscopic density $\left\langle n_{\mathbf{q}}\left(t\right)\right\rangle$ oscillates and reaches a maximum when the wavefunction is most localized, and a minimum when it is most delocalized. Each curve is colored as in Fig.~\ref{fig:KEandMicro}(b); note that the density oscillations of the white state are identical to the plotted black curve. Dashed lines show the averaged densities. (c) Maps of the interaction factor $\eta_{\mathbf{q}}$, equal to the time-averaged microscopic density (see text), for different polarizations. The colored dots mark the ground states after accounting for the interaction factor. Note that circular polarization retains the $D4$ symmetry of the single particle dispersion. 
	}
\end{figure}

\begin{figure} 
	\includegraphics[width=\columnwidth]{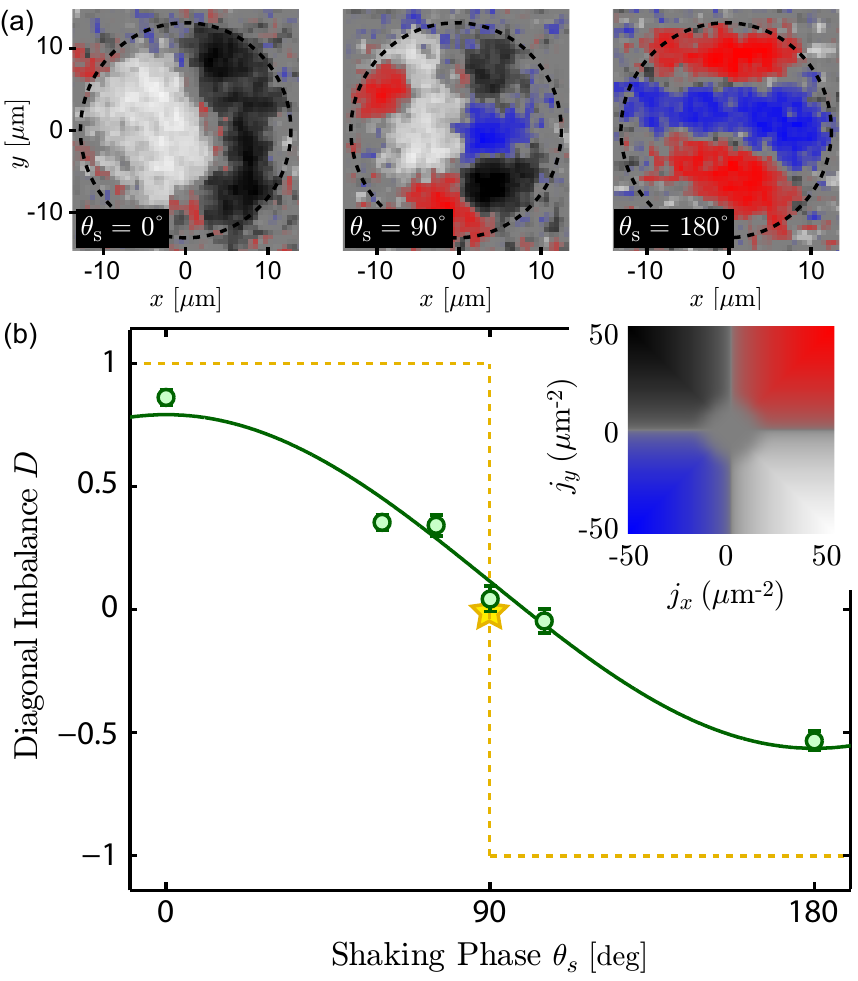}
	\caption{\label{fig:shakePhaseData}
		Observed coupling of interaction and momentum.
		(a) Example, reconstructed domain structures (see text) representing the density profiles of atoms in each well, measured after crossing the effectively ferromagnetic phase transition with the shaking polarizations indicated on each image. The dashed circles guide the eye to the region containing the condensate. The correspondence between color and pseudo-spin density (see text) is shown in the upper-right corner of panel (b).
		(b) The imbalance $D$ (see text) of the atomic populations between the two quasi-momentum diagonals characterizes the anisotropy which results from the quasimomentum-dependent interactions for different polarizations. The solid curve is a sinusoidal fit. The orange, dashed curve shows the expected imbalance in the absolute ground state; the star emphasizes that the expected imbalance is $D=0$ for circular shaking ($\theta_s=90^\circ$).
	}
\end{figure}

Our experiments utilize disk-shaped Bose-Einstein condensates of cesium atoms prepared
in a 2D, square optical lattice. The lattice depths along both directions are equal and small enough to maintain superfluidity of the gas. The lattice can then be shaken with identical peak-to-peak amplitudes $s$ and angular frequencies $\omega$ along both axes, see Fig.~\ref{fig:KEandMicro}(a). The shaking frequency is chosen to be slightly higher than the excitation gap at zero momentum in the lattice \cite{Parker2013}. See supplement for details \cite{supplement}.

When the shaking amplitude $s$ exceeds a critical value $s_c$, the single particle dispersion $E_{\mathrm{kin}}$ develops four minima at momenta $\mathbf{q}=(\pm q^{*},\pm q^{*})$ and $(\pm q^{*},\mp q^{*})$, where $q^*$ is controlled by $s$, see Fig.~\ref{fig:KEandMicro}(b). We calculate the effective dispersion of this periodically-modulated system using Floquet theory \cite{supplement}.
The four-fold degeneracy is the result of the $D4$ symmetry of the lattice, a 2D generalization of previous experiments in 1D \cite{Parker2013,Ha2015,Clark2016,Feng2017}. Similar to the 1D system, the change in dispersion induces a phase transition in which the condensate segregates into domains, each containing atoms occupying one of the four minima. Since the single particle Hamiltonian is separable along the lattice axes, the kinetic energy is independent of the shaking polarization $\theta_s$, defined as the relative phase between the two shaking lattices, see Fig.~\ref{fig:KEandMicro}(c).

We first explore the intriguing interplay between micromotion and interactions. Examples of the micromotion, the back-and-forth oscillation of the atomic wavefunction during one period $\tau$ of the lattice shaking, are shown in Fig.~\ref{fig:interactionsVsPolarization}(a). Since the atomic density depends on the wavefunction spread in both $x$-- and $y$--directions, interactions effectively couple the motion in the two directions and destroy the separability of the system. In particular, the micromotion creates a microscopic density enhancement factor $\braket{n_{\mathbf{q}}(t)}=d^2\int_0^d\int_0^d dxdy|\psi_{\mathbf{q}}(x,y,t)|^4$, where $\psi_{\mathbf{q}}(x,y,t)$ is the (unit-normalized) Floquet steady state wavefunction and the angle brackets denote the expectation value~\cite{supplement}. The enhancement factor characterizes the ratio of the average density in a lattice site to the coarse-grained density $\rho$.
This enhancement factor oscillates at the shaking frequency and can differ between the four kinetic energy minima, as shown in Fig. 2(b).
In this example, the wavefunction expands and contracts along the $x$-- and $y$--axes in-phase for momenta along the lattice shaking direction, leading to strong oscillations in density. In contrast, the wavefunctions along the $x$-- and $y$--axes oscillate out-of-phase for states with momentum perpendicular to the axis of lattice motion, reducing the density oscillation. For circular shaking the wavefunctions oscillate $90^\circ$ out of phase for all four momentum states, causing all four to have the same amplitude of density oscillation and therefore the same interaction energy.

Since the typical dynamics of the condensate, including the formation of domains after the phase transition, occur on timescales spanning many shaking periods, they are predominantly sensitive to the interaction energy, $\mathcal{E}_{\mathbf{q}}=\rho \overline{g(t)\braket{n_{\mathbf{q}}(t)}}$, where the bar denotes time-averaging over one shaking period. Therefore, we define the interaction factor,
\begin{equation}
\label{IntFactDef}
\eta_{\mathbf{q}}=\frac{1}{g_0} \overline{g(t)\braket{n_{\mathbf{q}}(t)}},
\end{equation}
which accounts for the interplay between the interaction strength and the micromotion, see Eq.~(\ref{Eq1}).

In the first stage of our experiments, with static interactions $g(t)=g_0$, we control the interaction-momentum coupling by tuning the shaking polarization, as shown in Fig.~\ref{fig:interactionsVsPolarization}(c). To leading order in $q/q_L$ the interaction factor is,
\begin{equation}\label{EqScheme1}
\eta_{\mathbf{q}}=\alpha + \beta s^{2} \cos\theta_s q_x q_y,
\end{equation}
where $\alpha$ and $\beta$ are dimensionless constants that depend on the shaken lattice parameters \cite{supplement}. The strength of this effect is greatest for linear shaking ($\theta_s=0^\circ$ or $180^\circ$), with which the momentum states along the axis of lattice motion experience much stronger density modulation, leading to a higher interaction factor than the momentum states perpendicular to the axis of lattice motion (hereafter ``off-diagonal states''), whose density is more consistent over time.
This effect causes domains to form preferentially in the off-diagonal states.

We test for the presence of interaction-momentum coupling by driving condensates across the phase transition with different shaking phases $\theta_s$ and measuring the resulting quasimomentum distribution. After loading the condensate into the lattice, we linearly ramp up the shaking amplitude, exceeding the critical amplitude and thus driving the condensate across the phase transition.
After a brief time-of-flight we measure the density distributions $n_i(\mathbf{r})$ of atoms occupying the quasimomentum state in the $i$'th quadrant; for example, $n_1$ is the density in the ($+q^*,\,+q^*$) state. Finally, we calculate the pseudo-spin density along each lattice axis, $j_x=n_1+n_4-n_2-n_3$ and  $j_y=n_1+n_2-n_3-n_4$. See supplement for details \cite{supplement}.

Typical reconstructed domain images for various shaking polarizations are shown in Fig.~3(a). To better quantify the biasing of the domains toward particular wells for ensembles of many images, we introduce an imbalance factor $D=(N_2+N_4-N_1-N_3)/N_{\mathrm{tot}}$, where $N_i$ is the population in the $i-$th quadrant and $N_{\mathrm{tot}}$ is the total atom number.
We observe a clear, polarization-dependent biasing of the domains toward forming in off-diagonal states, indicative of interaction-momentum coupling, see Fig.~\ref{fig:shakePhaseData}(b). For linear shaking, which maximizes the interaction-momentum coupling, the diagonal imbalance approaches 1 (-1) with $\theta_s=0^\circ$ ($180^\circ$), as expected.
Under these conditions, the $D4$ symmetry of the ground states is clearly broken by interactions. As the shaking polarization becomes more circular, the imbalance is progressively reduced.
For precisely circular shaking ($\theta_{s}=90^\circ$) the interaction-momentum coupling disappears and the $D4$ symmetry is restored, resulting in a diagonal imbalance of $D=0.04(5)$ consistent with zero.
Because of the finite ramp speed in our experiments, the phase transition is not adiabatic \cite{Clark2016}. As a result, the bias of the gas toward off-diagonal states increases with the energy difference between the wells. This effect causes the magnitude of the diagonal imbalance to smoothly increase as the interaction-momentum coupling is enhanced by tuning the shaking polarization from circular toward linear, as observed in Fig.~\ref{fig:shakePhaseData}(b).

\begin{figure} 
	\includegraphics[width=\columnwidth]{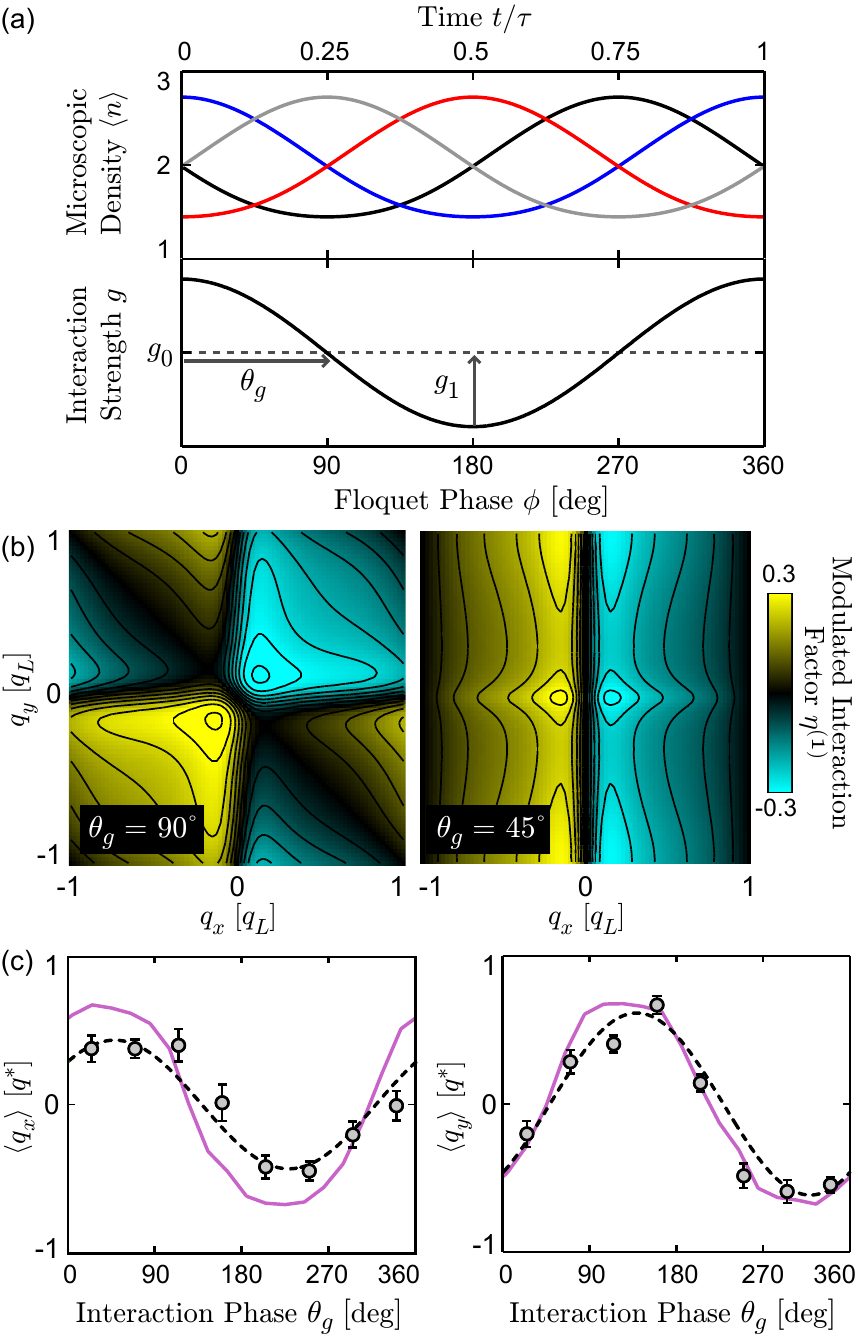}
	\caption{\label{fig:shakeAndScatMod} Density-dependent synthetic field from synchronized shaking and interaction strength modulation.
		(a) The upper panel plots the mean, microscopic density for circular shaking ($\theta_s=90^\circ$). Each curve is colored as in Fig.~\ref{fig:KEandMicro}(b).
		The lower panel shows the modulated interaction strength
		$g\left(t\right)=g_{0}-g_{1}\cos\left(\omega t - \theta_{g}\right)$. The modulated interactions raise (lower) the energy of quasimomentum states whose density oscillates in phase (out of phase) with the interaction modulation.
		(b) Modulated interaction factors for $\theta_{g}=90^{\circ}$ (left) and $\theta_{g}=45^{\circ}$ (right).
		(c) Measurement of the average quasimomentum of the condensate ($q^*=0.08~q_L$) in the presence of the interaction-induced field (circles). Error bars show standard error. The dashed curves show simultaneous, sinusoidal fits, which yield a phase offset of only $4\pm3^\circ$ from expectations. Simulations using the Gross-Pitaevskii equation \cite{Anderson2017} (solid magenta curves) agree well with the experiment.}
\end{figure}

In the second stage of our experiments, we generate a density-dependent gauge field by applying synchronized shaking and interaction strength modulation. We tune the magnetic field near a Feshbach resonance \cite{Chin2010} to modulate the interaction strength as $g(t)=g_0-g_1\cos\left(\omega t-\theta_{g}\right)$ at the same frequency as the lattice shaking and with phase $\theta_{g}$, see Fig.~\ref{fig:shakeAndScatMod}(a).
In this case, the interaction-momentum coupling can be understood intuitively by comparing the microscopic density and the interaction strength during each shaking period, see Fig.~\ref{fig:shakeAndScatMod}(a). When the interaction strength oscillates in phase (out of phase) with the density, the interaction energy is maximized (minimized).

To quantify the interaction-induced field, the interaction factor can be decomposed as, see Eq.~(\ref{IntFactDef}),
\begin{equation}
\eta_{\mathbf{q}}=\eta_{\mathbf{q}}^{(0)}+\frac{g_1}{g_0}\eta_{\mathbf{q}}^{(1)},
\end{equation}
where $\eta_{\mathbf{q}}^{(0)}=\overline{\braket{\eta_{\mathbf{q}}(t)}}$ is the static interaction factor and $\eta_{\mathbf{q}}^{(1)}=-\overline{\braket{\eta_{\mathbf{q}}(t)}\cos(\omega t -\theta_g)}$ is the modulated interaction factor.
We use circular shaking ($\theta_s=90^\circ$) so that the static interaction factor maintains the $D4$ symmetry.
For small momentum $|\mathbf{q}|\ll q_L$ the modulated interaction factor takes the form \cite{supplement},
\begin{equation}\label{Eq:Scheme2}
\eta_{\mathbf{q}}^{(1)}=-\sqrt{\frac{\alpha\beta}{2}}\, s\, \mathbf{e}_{\Theta}\cdot\mathbf{q},
\end{equation}
which corresponds to the density-dependent gauge potential, 
\begin{equation}
\label{Eq:GaugePotential}
\mathbf{A}(\rho)=\sqrt{\frac{\alpha\beta}{2}}\,ms\,g_1\rho\,\mathbf{e}_{\Theta},
\end{equation}
whose direction is given by $\mathbf{e}_{\Theta}$ with $\Theta\equiv\theta_g-\theta_s/2$. The equivalent treatment of the gauge field in terms of an occupation-dependent Peierls phase does not rely on the small momentum limit \cite{supplement}. The static interaction factor $\eta_{\mathbf{q}}^{(0)}$, which does not correspond to a gauge potential, can be made negligible by reducing the average interaction strength $g_0$. Salient examples of the modulated interaction factors from a numerical calculation are shown in Fig.~\ref{fig:shakeAndScatMod}(b).

Experimentally, we test for the interaction-induced gauge field by measuring the bias toward particular quasimomenta as a function of the interaction phase $\theta_g$. We first prepare the condensate in a stationary lattice with an oscillating scattering length. We then begin to circularly shake the lattice, linearly increasing the shaking amplitude and driving the system across the phase transition.
After a brief settling time, we measure the momentum distribution $\rho(\mathbf{q})$ based on time-of-flight expansion \cite{Feng2017} and calculate the average quasimomentum $\braket{\mathbf{q}}=\int\mathbf{dq}\,\mathbf{q}\rho(\mathbf{q})$ \cite{supplement}.

The average quasimomentum after the phase transition shows a clear bias depending on the interaction modulation phase $\theta_g$, indicative of the interaction-induced gauge field, see Fig.~\ref{fig:shakeAndScatMod}(c). Based on the form of the gauge potential shown in Eq.~(\ref{Eq:Scheme2}), we expect the biasing along the $x-$ and $y-$axes to take the approximate forms $\braket{q_x}\propto \cos(\theta_g-45^\circ)$ and $\braket{q_y}\propto \sin(\theta_g-45^\circ)$. Simultaneous, sinusoidal fits to the data in Fig.~\ref{fig:shakeAndScatMod}(c) yield a phase consistent with this prediction. The magnitude of the bias in momentum does not reach $q^*$, since it depends sensitively on the dynamics of crossing the phase transition \cite{Parker2013, Clark2016} as well as the magnitude of the gauge potential.
In principle, the size of the interaction induced field, and therefore the bias, can be increased by using a larger interaction modulation amplitude. However, doing so can induce other instabilities in the gas \cite{Pollack2010,Clark2017,Feng2018}. 

To confirm that the magnitude of the observed effect matches theoretical expectations, we have performed simulations of this experiment using the Gross-Pitaevskii equation \cite{Anderson2017}. The resulting magenta curves in Fig.~\ref{fig:shakeAndScatMod}(c), which show the average outcomes of five simulations at each $\theta_g$ ($20^\circ$ steps) with different random noise seeds, agree nicely with our experiments.

In summary, we have demonstrated an interaction-induced gauge field based on synchronous lattice shaking and interaction strength modulation. Our work presents a paradigm to guide the simulation of gauge field theories using ultracold atom systems. For example, this scheme can be used directly to simulate the anyon-Hubbard model \cite{Keilmann2011,Greschner2015,Cardarelli2016,Strater2016}, as detailed in the supplement \cite{supplement}.

\begin{acknowledgments}
	\textit{Acknowledgments.} L.W.C. was supported by a Grainger Graduate Fellowship.
	This work was supported by NSF grant PHY-1511696, Army Research Office-Multidisciplinary Research Initiative grant W911NF-14-1-0003, and the University of Chicago Materials Research Science and Engineering Center, which is funded by the National Science Foundation under award number DMR-1420709.
\end{acknowledgments}

\setcounter{equation}{0}
\renewcommand{\theequation}{S\arabic{equation}}

\section{Supplementary Material}

In the first section of the supplement we provide details on the experimental procedures used to test for the density-dependent gauge field. In the remainder of the supplement we sketch the theoretical treatment which
demonstrates how our synthetic gauge potentials arise. In the second section, we treat our system in momentum-space within the mean-field approximation, deriving the key results presented in the main text. In the third section, we lift the mean-field approximation and develop a tight-binding model for a one-dimensional version of our system, showing that synchronized shaking and interaction strength modulation lead to a density-dependent Peierls phase for tunneling in the lattice. Finally, using this tight-binding picture, we present a mapping of our bosons with an interaction-induced gauge field onto the Anyon-Hubbard model for particles with fractional exchange statistics propagating in a one-dimensional chain. 

\section{Experiment Details}

Our experiments utilize Bose-Einstein condensates of $N=30,000$ cesium atoms prepared
in a harmonic trap with horizontal frequencies $\omega_{x}\approx\omega_{y}=2\pi\times8$Hz and tight vertical confinement of
$\omega_{z}=2\pi\times200$~Hz. We load the atoms into a 2D, square optical lattice with lattice spacing $d=\pi /q_{L}=532$~nm. The lattice depths $V_{0}$ along both directions are equal and small enough to maintain superfluidity of the gas. The lattice can then be shaken with identical peak-to-peak amplitudes $s$ and angular frequencies $\omega$ along both axes.

To test for the presence of interaction-momentum coupling in the presence of the shaken lattice but without interaction strength modulation, we drive condensates across the phase transition with different shaking phases $\theta_s$ and measured the resulting quasimomentum distribution, as shown in Fig.~3 of the main text. 
For these experiments, we use a lattice depth of $V_0=8.86$~$E_\mathrm{R}$, where $E_{{\rm R}}=\hbar^{2}q_{L}^{2}/2m=h\times 1.33$~kHz is the recoil energy, and shaking frequency $\omega=2\pi\times8$~kHz; at this depth, the bare tunneling energy is $J=h\times33$~Hz. We employ a constant scattering length of $a_0=16~a_B$, where $a_B$ is the Bohr radius, such that $g_0\rho=h\times80$~Hz in the center of the gas. After loading the condensate into the lattice, we linearly increase the shaking amplitude to $s=20$~nm over 100~ms, exceeding the critical amplitude $s_c=13$~nm to drive the condensate across the phase transition. This slow ramp causes domains to form close to the critical point, where the four kinetic energy wells are shallow and therefore the relative importance of the interaction-momentum coupling energy to determining the domain structure is enhanced. We subsequently increase the shaking amplitude to $s=32$~nm over 10~ms in order to increase the momentum separation between the states, making them easier to distinguish during detection. Finally, we hold the gas for another 120~ms to ensure that domains have clearly formed, after which we perform a short (5~ms) time-of-flight which enables us to reconstruct the original, \textit{in-situ} domain distribution; see Ref.~\cite{Clark2016} for details on the reconstruction procedure. In particular, we can extract the density distributions $n_i(\mathbf{r})$ of atoms occupying the quasimomentum state in the $i$'th quadrant; for example, $n_1$ is the density in the ($+q^*,\,+q^*$) state. From these we calculate the pseudo-spin density along each lattice axis, $j_x=n_1+n_4-n_2-n_3$ and  $j_y=n_1+n_2-n_3-n_4$. 

To test for the density-dependent gauge field we apply synchronized lattice shaking and interaction strength modulation and measure the bias toward particular quasimomenta as a function of the interaction phase $\theta_g$. Here, we prepare the condensate in a stationary lattice of depth $V_0=4$~$E_\mathrm{R}$ with a static scattering length of $a_0=16~a_B$ before ramping up the scattering length modulation amplitude to a maximum value of $a_{1}=25~a_B$ over 25~ms. At this stage, the modulated interaction energy scale is $g_1\rho=h\times125$~Hz, comparable to the bare tunnel coupling of $J=h\times113$~Hz. We then begin to circularly shake the lattice with frequency $\omega=2\pi\times6.3$~kHz, increasing the shaking amplitude to $s=26$~nm over 70~ms, which drives the system across the phase transition. After a settling time of 10~ms, we measure the momentum distribution $\rho(\mathbf{q})$ based on time-of-flight expansion \cite{Feng2017} and calculate the average quasimomentum $\braket{\mathbf{q}}=\int\mathbf{dq}\mathbf{q}\rho(\mathbf{q})$. Note that experiments performed for a wide range of lattice depths and shaking frequencies exhibit qualitatively similar results. 

\section{Interaction-Induced Gauge Fields from Floquet Theory}

We describe our system, consisting of a Bose condensate in a shaken two-dimensional
optical lattice with the many-body Hamiltonian
\begin{eqnarray}
\mathcal{H}\left(t\right) & = & \int d\mathbf{r}\,\hat{\psi}^{\dagger}\left(\mathbf{r},t\right)\left(H_{0}\left(t\right)-\mu\right)\hat{\psi}\left(\mathbf{r},t\right)\label{eq:H-many-body}\\
& + & \frac{g\left(t\right)}{2}\int d\mathbf{r}\,\hat{\psi}^{\dagger}\left(\mathbf{r},t\right)\hat{\psi}^{\dagger}\left(\mathbf{r},t\right)\hat{\psi}\left(\mathbf{r},t\right)\hat{\psi}\left(\mathbf{r},t\right),\nonumber
\end{eqnarray}
where $\hat{\psi}^{\dagger}\left(\mathbf{r},t\right)$ $\left(\hat{\psi}\left(\mathbf{r},t\right)\right)$
creates (destroys) a boson at position $\mathbf{r}=\left(x,y\right)$
and time $t$, $H_{0}\left(t\right)=-\frac{\hbar^2}{2m}\nabla^{2}+V_{L}\left(\mathbf{r}-\delta \mathbf{r}\left(t\right)\right)$
is the time-dependent single-particle Hamiltonian in a shaken lattice,
$\mu$ is the chemical potential of the Bose gas, and the interaction
constant $g\left(t\right)$ is periodically modulated.

Central for this paper is the assumption that the interaction energy is sufficiently
small so that the system samples only the lowest Floquet band.
This
assumption is well justified for the experimental parameters considered.
While the question of optimal preparation of many-body states is still
a generally open question, we assume we can use the Floquet adiabatic
preparation scheme.

\paragraph{Single particle Floquet.\textendash{}}

Our lattice $V_{L}\left(\mathbf{r}\right)=V_{L}\left(x\right)+V_{L}\left(y\right)$ is separable,
with a shaking function $\delta \mathbf{r} \left(t\right)=s\left[\sin\left(\omega t-\phi_x \right),\sin\left(\omega t-\phi_y\right)\right]/2$
composed of a pure sine-wave oscillating at frequency $\omega=2\pi/\tau$;
the phases $\phi_x=-\phi_y=-\theta_{s}/2$ define the shaking polarization $\theta_{s}$, which tunes the system from circular ($\theta_{s}=\left(j+1/2\right)\pi$ where $j=0,1,...$ is an integer) to linear shaking ($\theta_{s}=j\pi$). 
Separability of the lattice transfers to the single particle Floquet band
structure, since $H_{0}\left(t\right)=H_{x}\left(t\right)+H_{y}\left(t\right)$,
where $H_{x}\left(t\right)$ and $H_{y}\left(t^{\prime}\right)$ commute
at distinct times. The Floquet Hamiltonian $H^{{\rm floq}}=-\frac{i}{T}\ln\mathcal{U}\left(\tau\right)$
is therefore the sum of the Floquet Hamiltonian as calculated along
each direction individually, i.e., $H^{{\rm floq}}=H_{x}^{{\rm floq}}+H_{y}^{{\rm floq}}$,
with Floquet eigenstates given by product states $\psi_{{\rm 2D}}^{{\rm floq}}=\psi_{x}^{{\rm floq}}\psi_{y}^{{\rm floq}}$,
with corresponding energy $E_{\mathbf{q}}^{{\rm floq}}=E_{q_{x}}^{{\rm floq}}+E_{q_{y}}^{{\rm floq}}$.

We chose our shaking frequency
$\hbar \omega=E_{{\rm sp}}+\delta$ to be near resonant with the
zero-momentum band gap $E_{{\rm sp}}$ of the two lowest bands, and
choose $\delta>0$. This allows us to approximate $H_{x,y}^{{\rm floq}}$
along direction $x,y$ by a two-band model \cite{Zheng2014}:
\begin{equation}
H_{x,y}^{{\rm floq}}\left(q_{x,y}\right)=\left(\begin{array}{cc}
E^{s}\left(q_{x,y}\right) & e^{\pm i\phi_{x,y}}\Omega_{q_{x,y}}\\
e^{\mp i\phi_{x,y}}\Omega_{q_{x,y}}^{*} & E^{p}\left(q_{x,y}\right)-\omega
\end{array}\right)\label{eq:H-two-band}.
\end{equation}
The coupling $\Omega_{q_{x,y}}\sim s$ can be used to drive the single particle
dispersion along each axis from a single well at $q_{x,y}=0$ to a double-well structure
at $q_{x,y}=\pm q^{*}$ at the critical shaking amplitude $s=s_{c}$.

Diagonalizing the $2 \times 2 $ Hamiltonian in Eq.~(\ref{eq:H-two-band}) gives two
Floquet bands. We consider only the band adiabatically connected 
to the $s$ band in the limit of zero shaking. 
Since the rotating-wave approximation is valid, $\psi_{x,y}^{{\rm floq}}$
is composed of only an $s$-band term coupled to a $p$-band term
rotating as $e^{i\omega t}$;
we denote the Floquet
wavefunction for this state as 
$\psi_{q}^{{\rm floq}}\left(x,t\right)=c_{q}^{\left(s\right)}u_{q}^{\left(s\right)}\left(x\right)+e^{i\left(\omega t-\phi\right)}c_{q}^{\left(p\right)}u_{q}^{\left(p\right)}\left(x\right)$, 
where $u_{q}^{(s,p)}$ and 
$c_{q}^{(s,p)}$ are respectively the Bloch eigenfunction and Floquet
coefficient for the $s,p$ bands in question.
For notational convenience, we will  henceforth drop the superscript $\textrm{floq}$ on Floquet states. We will also sometimes drop the subscript on $q_{x,y}$ and $\phi_{x,y}$; this is meant to imply that there are two equations, one for $\{x, q_x, \phi_x\}$, and one for $\{y, q_y, \phi_y\}$.

We conclude with a key observation: because of the separability of
$H_{0}\left(t\right)$, the relative shaking phase between the $x$
and $y$ components will not enter in the Floquet energy.  

\paragraph{Effects of time dependent interactions.\textendash{}}

We now consider a Bose condensate, and turn to the effects of many body interactions through
the mean field interaction energy, $\mathcal{E}_{I}\left(t\right)=\frac{g\left(t\right)}{2}\int n^{2}\left(\mathbf{r},t\right)d\mathbf{r}$.
Ultracold bosons will tend to occupy the combination of the four kinetic energy minima which also minimizes
the total, time-averaged interaction energy $\bar{\mathcal{E}_{I}}=\frac{1}{\tau}\int_{0}^{\tau}\mathcal{E}_{I}\left(t\right)dt$
over one period $\tau=2\pi/\omega$. 

Since the interaction energy is minimized when all of the atoms occupy the same quasi-momentum state, we consider the marginal interaction energy for a particular quasi-momentum state in the ground Floquet band,
\begin{eqnarray}
\mathcal{E}_{\mathbf{q}}\left(t\right)  =  \frac{\partial\mathcal{E}_{I}}{\partial N}  =  \rho g\left(t\right)\left\langle n_{\mathbf{q}}\left(t\right)\right\rangle.
\end{eqnarray}
The factor,
\begin{eqnarray}
\left\langle n_{\mathbf{q}}\left(t\right)\right\rangle = d^2 \int_0^d\int_0^d\left|\psi_{\mathbf{q}}\left(x,y,t\right)\right|^{4}dxdy,
\end{eqnarray} 
is the density enhancement factor that characterizes the increase in interaction energy due to the microscopic density modulation induced by the lattice structure. Here, the Floquet wavefunctions are normalized such that $\int_0^d\int_0^d\left|\psi_{\mathbf{q}}\left(x,y,t\right)\right|^{2}dxdy=1$. Without any lattice, the density enhancement factor would take its minimum value $\left\langle n_{\mathbf{q}}\left(t\right)\right\rangle=1$. In a shaking lattice, the density enhancement factor becomes greater than one and oscillates at the shaking frequency $\omega$. 

The mean interaction energy per particle is,
\begin{equation}
\mathcal{E}_{\mathbf{q}}\equiv\overline{\mathcal{E}_{\mathbf{q}}\left(t\right)},
\end{equation}
where $\overline{f\left(t\right)}=\frac{1}{\tau}\int_{0}^{\tau}f\left(t\right)dt$
represents period averaging. We factor the energy as,
\begin{equation}
\mathcal{E}_{\mathbf{q}}=g_{0}\rho\eta_{\mathbf{q}},
\end{equation}
where
\begin{equation}
\eta_{\mathbf{q}}=\frac{1}{g_{0}}\overline{g\left(t\right)\left\langle n_{\mathbf{q}}\left(t\right)\right\rangle }
\end{equation}
is the interaction factor discussed in the main text. With $g\left(t\right)=g_{0}-g_{1}\cos\left(\omega t-\theta_{g}\right)$,
we see that $\eta_{\mathbf{q}}=\eta_{\mathbf{q}}^{\left(0\right)}+\frac{g_{1}}{g_{0}}\eta_{\mathbf{q}}^{\left(1\right)}$
naturally decomposes into a static term $\eta_{\mathbf{q}}^{\left(0\right)}=\overline{\left\langle n_{\mathbf{q}}\left(t\right)\right\rangle }$
and a dynamic term $\eta_{\mathbf{q}}^{\left(1\right)}=-\overline{\cos\left(\omega t-\theta_{g}\right)\left\langle n_{\mathbf{q}}\left(t\right)\right\rangle }$.

We now calculate the interaction factor, starting with the factorizeable
Floquet wavefunction $\psi_{\mathbf{q}}\left(x,y,t\right)=\psi_{q_{x}}\left(x,t\right)\psi_{q_{y}}\left(y,t\right)$,
where $\psi_{q}\left(x,t\right)=c_{q}^{\left(s\right)}u_{q}^{\left(s\right)}\left(x\right)+e^{i\left(\omega t-\phi\right)}c_{q}^{\left(p\right)}u_{q}^{\left(p\right)}\left(x\right)$;
factorizability implies
\begin{equation}
\left\langle n_{\mathbf{q}}\left(t\right)\right\rangle =\left\langle n_{q_{x}}\left(t\right)\right\rangle \left\langle n_{q_{y}}\left(t\right)\right\rangle.
\end{equation}
The density can then be expressed as
\begin{eqnarray}
n_q\left(x,t\right)=\left|\psi_q\left(x,t\right)\right|^{2} & = & n_{q}\left(x\right)+\delta n_{q}\left(x,t\right),
\end{eqnarray}
where  
\begin{eqnarray}
n_{q}\left(x\right) & = & \left|c_{q}^{\left(s\right)}u_{q}^{\left(s\right)}\left(x\right)\right|^{2}+\left|c_{q}^{\left(p\right)}u_{q}^{\left(p\right)}\left(x\right)\right|^{2},
\end{eqnarray}
and
\begin{eqnarray}
\delta n_{q}\left(x,t\right)
& = & \delta n_{q}\left(x\right) \cos\left(\omega t - \phi \right) \\
& = & 2\Re\left[c_{q}^{\left(s\right)}u_{q}^{\left(s\right)}\left(x\right)\left(c_{q}^{\left(p\right)}u_{q}^{\left(p\right)}\left(x\right)\right)^{*} e^{-i(\omega t -\phi)}\right], \nonumber
\end{eqnarray}
so
\begin{eqnarray}
\left\langle n_{q}\left(t\right)\right\rangle 
& = & \left\langle n_{q}\right\rangle + \left\langle \delta n_{q}\right\rangle \sin\left(\omega t-\phi\right)+\mathcal{O}\left(\delta n_{q}^{2}\right) \nonumber.
\end{eqnarray}

When calculating the static interaction factor,
\begin{equation}
\eta_{\mathbf{q}}^{\left(0\right)}=\overline{\left\langle n_{\mathbf{q}}\left(t\right)\right\rangle }=\overline{\left\langle n_{q_{x}}\left(t\right)\right\rangle \left\langle n_{q_{y}}\left(t\right)\right\rangle }
\end{equation}
we get
\begin{eqnarray}
\eta_{\mathbf{q}}^{\left(0\right)} & = & \left\langle n_{q_{x}}\right\rangle \left\langle n_{q_{y}}\right\rangle +\left\langle \delta n_{q_{x}}\right\rangle \left\langle \delta n_{q_{y}}\right\rangle \overline{\sin\left(\omega t-\phi_{x}\right)\sin\left(\omega t-\phi_{y}\right)} \nonumber\\
& = & \left\langle n_{q_{x}}\right\rangle \left\langle n_{q_{y}}\right\rangle +\left\langle \delta n_{q_{x}}\right\rangle \left\langle \delta n_{q_{y}}\right\rangle \cos\left(\theta_{s}\right)/2,
\end{eqnarray}
where $\phi_{x}-\phi_{y}=-\theta_{s}/2$, plus terms that are $\mathcal{O}\left(\delta n_{q}^{2}\right)$,
but which are isotropic to $\mathcal{O}\left(\mathbf{q}^{2}\right)$,
and therefore are not the dominant symmetry breaking terms.

Performing a small $\mathbf{q}$ expansion, $\left\langle \delta n_{q}\right\rangle \rightarrow\sqrt{2\beta}sq$,
and defining $\alpha=\left\langle n_{q_{x}}\right\rangle \left\langle n_{q_{y}}\right\rangle $,
we obtain Eq.~(4) in the main text:
\begin{equation}
\eta_{\mathbf{q}}^{\left(0\right)}=\alpha+\beta s^{2}q_{x}q_{y}\cos\theta_{s}
\end{equation}
where $\alpha$, and $\beta$ depend on the shaken lattice parameters. 
Numerical calculation of the interaction factor starting with the the full Floquet-Bloch approximation results in Fig. 2(c) of the main text.

\begin{widetext}
	
	Finally, the dynamic interaction factor $\eta_{\mathbf{q}}^{\left(1\right)}$,
	is given by
	\begin{eqnarray}
	\eta_{\mathbf{q}}^{\left(1\right)} & = & -\left\langle n_{q_{x}}\right\rangle \left\langle \delta n_{q_{y}}\right\rangle \overline{\cos\left(\omega t-\theta_{g}\right)\sin\left(\omega t-\phi_{y}\right)}-\left\langle n_{q_{y}}\right\rangle \left\langle \delta n_{q_{x}}\right\rangle \overline{\cos\left(\omega t-\theta_{g}\right)\sin\left(\omega t-\phi_{x}\right)}\\
	& = & -\left\langle n_{q_{x}}\right\rangle \left\langle \delta n_{q_{y}}\right\rangle \sin\left(\Theta\right)/2-\left\langle n_{q_{y}}\right\rangle \left\langle \delta n_{q_{x}}\right\rangle \sin\left(\Theta+\theta_{s}\right)/2
	+ \mathcal{O}(\mathbf{q}^2) \nonumber
	\end{eqnarray}
	Expanding in small momentum, and taking $\theta_{s}=90^{\circ}$,
	we have
	\begin{eqnarray}
	\eta_{\mathbf{q}}^{\left(1\right)} & = & -\sqrt{\alpha\beta/2}s\left(q_{x}\cos\Theta+q_{y}\sin\Theta\right) \\
	& = & -\sqrt{\alpha\beta/2}s\mathbf{q}\cdot\mathbf{e}_{\Theta} \nonumber
	\end{eqnarray}
	This analysis has yielded Eq.~(6) in the main text. Again,
	a complete numerical calculation of the interaction factor can be performed, resulting in Fig. 4(b) of the main text.
\end{widetext}

The energy shift due to the dynamic interaction factor can be understood as an interaction-induced synthetic gauge field by noting that a charged particle of mass $m$ in a gauge potential $\mathbf{A}$ experiences a momentum-dependent energy shift,
\begin{eqnarray}
E=-\frac{\mathbf{q}\cdot\mathbf{A}}{m},
\end{eqnarray}
where we have incorporated the hypothetical charge of the particle in the gauge potential itself. Equating this form with the mean interaction energy per particle from the dynamic interaction factor, $\mathcal{E}_\mathbf{q}=-\sqrt{\frac{\alpha\beta}{2}}s\rho g_1 \mathbf{q}\cdot\mathbf{e}_\Theta $, yields the interaction-induced synthetic gauge field,
\begin{eqnarray}
\mathbf{A}(\rho)=\sqrt{\frac{\alpha\beta}{2}}\,ms\,g_1\rho\,\mathbf{e}_{\Theta}
\end{eqnarray}
presented as Eq.~(7) of the main text.

\section{Tight-Binding Model with Density-Dependent Peierls Phase}	

In this section we derive a Hubbard-type model for our system in the tight-binding
limit with an occupation-dependent Peierls phase. We limit the discussion to one-dimension for clarity, but the extension to
multiple dimensions is straightforward. We will later show that, in
the mean-field and small-q limits taken above, this approach reproduces
our previous results. However, the tight-binding model provides a more complete picture of the interaction-induced gauge field produced
in our system. Moreover, in the next section we utilize this tight-binding
model to show that our system can be mapped onto the Anyon-Hubbard
model for anyons in a one-dimensional lattice.

In the tight-binding limit of a deep lattice, the unshaken $j$'th Bloch wavefunction
can be written simply as a sum of $j$'th harmonic oscillator eigenstates
$\phi^{(j)}(x)$ centered on each lattice site, 
\begin{equation}
u_{q}^{(j)}=\sum_{n}e^{iqdn}\phi^{(j)}(x-dn).
\end{equation}
As before, we assume
that the interaction energy is sufficiently small that the system
samples only the lowest Floquet band. For small shaking amplitudes,
the lowest Floquet wavefunctions are predominantly composed of the
$s$-state with a small admixture $\epsilon\frac{s}{d}$ ($\epsilon\ll1$)
of the $p$-state,

\begin{equation}
\psi_{q}(x,t)=u_{q}^{(s)}+\epsilon\frac{s}{d}e^{i(\omega t-\phi)}u_{q}^{(p)},
\end{equation}
where we have assumed that the detuning $\delta(q)\equiv\omega-(E^{p}(q)-E^{s}(q))$
is large compared to the bandwidth, such that it can be treated as
essentially constant across the Brillouin zone.

Here, we wish to work with the corresponding Floquet
Wannier functions on each site, 

\begin{equation}
w_{n}(x,t)=\phi^{(s)}(x-dn)+\epsilon\frac{s}{d}e^{-i(\omega t-\phi)}\phi_{q}^{(p)}(x-dn).
\end{equation}
Having restricted to the lowest Floquet band, we can write the field
operator as

\begin{equation}
\hat{\psi}(x)=\sum_{n}w_{n}(x)b_{n}
\end{equation}
where the operator $b_n$ annihilates a boson in the
lowest Floquet band at site $n$. Under these assumptions, the effective
Hamiltonian (see Eq. S1) becomes

\begin{equation}
\mathcal{H}_{eff}=-J\sum_{n}\left(b_{n}^{\dagger}b_{n+1}+b_{n+1}^{\dagger}b_{n}\right)+\frac{1}{2}\sum_{npqr}U_{npqr}b_{n}^{\dagger}b_{p}^{\dagger}b_{q}b_{r}
\end{equation}
where the bare tunneling $J$ is determined by the width of the lowest
Floquet kinetic energy band, the interaction coefficients are
\begin{equation}
U_{npqr}=\frac{1}{T}\int_{0}^{T}\mathrm{dt}g(t)\int\mathrm{dx}w_{n}^{*}(x,t)w_{p}^{*}(x,t)w_{q}(x,t)w_{r}(x,t),
\end{equation}
and we have assumed an infinite chain of sites for simplicity. 

We can greatly simplify the interaction terms by noting that, just
as the tight-binding limit allows us to neglect tunneling terms beyond
nearest-neighbors, the wannier functions are sufficiently well localized
that we can drop most of the terms in the series. As usual, the dominant
term arises from the on-site interaction term $U_{nnnn}$. However,
here the next-lowest order terms, for example $U_{nnnn+1}$ in which
a single neighboring site is included, are also relevant. These terms
encode tunneling processes whose amplitude depends on the number of
atoms occupying the sites tunneled between, which give rise to the
interaction-induced gauge field that we have observed. Dropping all
of the higher order terms, the Hamiltonian becomes,
\begin{multline}
\mathcal{H}_{eff}=-J\sum_{n}\left(b_{n}^{\dagger}b_{n+1}+\mathrm{H.c.}\right)+\frac{1}{2}\sum_{n}U_{0}N_{n}^{2}+\\ 
\sum_{n}\left(b_{n}^{\dagger}\left(U_{1}N_{n}+U_{-1}^{*}N_{n+1}\right)b_{n+1}+ \mathrm{H.c.}\right) 	
\end{multline}
where H.c. denotes the Hermitian conjugate of the previous term, $N_{n}\equiv b_{n}^{\dagger}b_{n}$ is the atom number in site $n$, and the interaction coefficients
$U_{0}\equiv\frac{1}{T}\int_{0}^{T}\mathrm{dt}g(t)\int\mathrm{dx}\left|w_{0}(x,t)\right|^{4}$,
$U_{1}\equiv\frac{1}{T}\int_{0}^{T}\mathrm{dt}g(t)\int\mathrm{dx}\left|w_{0}(x,t)\right|^{2}w_{0}^{*}(x,t)w_{1}(x,t)$,
and $U_{-1}\equiv\frac{1}{T}\int_{0}^{T}\mathrm{dt}g(t)\int\mathrm{dx}\left|w_{0}(x,t)\right|^{2}w_{0}^{*}(x,t)w_{-1}(x,t)$
are determined by the shaken lattice and interaction modulation parameters.

To show that this result corresponds to an interaction-induced gauge
field we rewrite the effective Hamiltonian in the form,

\begin{multline}
\mathcal{H}_{eff}=-\sum_{n}\left(b_{n}^{\dagger}\tilde{J}_{n,n+1}b_{n+1}+\mathrm{H.c}\right)+\\\frac{1}{2}\sum_{n}U_{0}N_{n}(N_{n}-1),
\label{Heff}
\end{multline}
which appears as a Bose-Hubbard model but with an occupation-dependent
tunneling amplitude

\begin{equation}
\tilde{J}_{n,n+1}\equiv J-U_{1}N_{n}-U_{-1}^{*}N_{n+1}.
\label{eqn:JfromU}
\end{equation}
When the coefficient
$U_{1}$ is complex, this results in an occupation-dependent Peierls
phase for the tunneling amplitudes, which is equivalent to the interaction-induced
gauge field \cite{Keilmann2011,Greschner2014,Greschner2015}. For clarity, in the remainder of this section, we will assume that the average
interaction strength $g_{0}$ is negligible and focus
on the gauge field which arises from $g_{1}$. Evaluating the integrals
above, we find the expressions for $U_{\pm1}$:

\begin{equation}
U_{\pm1}=\pm\frac{g_{1}s\gamma}{2}e^{-i(\theta_{g}-\phi)},
\label{eqn:Upm1}
\end{equation}
where the constant $\gamma$ is determined by the shaken lattice parameters.

To verify that this result is equivalent to the mean-field, momentum-space
result above, we can take the mean-field approximation where
$N_{n}=N_{n+1}=\rho d^{2}$. In that case, the mean-field effective
tunneling is $\tilde{J}_{MF}=J+isg_{1}\rho\gamma d^{2}\cos(\text{\ensuremath{\Theta}})$.
Transforming back to momentum-space with momentum $q$, the ordinary
tunneling term $J$ contributes the cosine band $E_{q}=-2J\cos(qd)$,
while the imaginary density-dependent term in $\tilde{J}_{MF}$ adds a sinusoidal contribution,
\begin{equation}
E_{q_x,MF}=-\sqrt{\alpha\beta/2}\frac{s}{md}\cos(\Theta)\sin(q_xd)
\end{equation}
where we have identified $g_{1}\rho\gamma d^{3}m\equiv\sqrt{\alpha\beta/2}$
by comparing our two derivations (in both cases, the additional constants
$\alpha,\beta,$ and $\gamma$ are determined by the lattice parameters
). In the limit $q_xd\ll1$, where $\sin(q_xd)\approx q_xd$, this expression
reproduces the $x$-component of Eq.~(S17) above for the mean-field
momentum-dependent energy shift from the interaction-induced gauge
potential. This correspondence verifies that the real-space and
momentum-space approaches are equivalent. Moreover, this derivation
provides the complete form of the energy shift across the Brillouin
zone, showing that it is sinusoidal in quasimomentum.

Returning to the full treatment without the mean-field approximation, we can substitute Eq.~(\ref{eqn:Upm1}) into Eq.~(\ref{eqn:JfromU}) to find
that the effective tunneling takes the simple form, 
\begin{equation}
\tilde{J}_{n,n+1}\equiv J+iJ_{int}\cos(\Theta)\left(N_{n}+N_{n+1}\right),
\end{equation}
where the real constant $J_{int}\equiv\frac{g_{1}s\gamma}{2}$ is
determined by the shaken lattice and interaction-strength modulation
parameters. The tunneling coefficient can be re-written in the form, 
\begin{equation}
\tilde{J}_{n,n+1}\equiv J_{n,n+1}\exp\left(i\theta_{n,n+1}\right),
\end{equation}
with magnitude,
\begin{equation}
J_{n,n+1}\equiv \sqrt{J^2+J_{int}^2\cos^2(\Theta)\left(N_n+N_{n+1}\right)^2},
\end{equation}
and an occupation-dependent Peierls phase,
\begin{equation}
\theta_{n,n+1}\equiv \arctan\left(\frac{J_{int}\cos(\Theta)(N_n+N_{n+1})}{J}\right).
\label{eqn:PeierlsPhase}
\end{equation}
This occupation-dependent Peierls phase encodes the interaction-induced gauge potential which is experimentally demonstrated in our work. 

In some cases systems with apparent gauge fields can be directly mapped to a trivial Hamiltonian with no gauge field; for an example, see the supplementary material of Ref.~\cite{Greschner2014}. We note that, unlike the trivial case explained there, our Peierls phase depends on the \textit{sum} of the occupations on adjacent sites (Eq.~\ref{eqn:PeierlsPhase}), rather than the difference. Moreover, in the next section we present a mapping of our system onto the Anyon-Hubbard model \cite{Keilmann2011,Greschner2015,Cardarelli2016,Strater2016}, precluding the possibility that it can generally be mapped onto an ordinary Bose-Hubbard model.

\section{Realizing the One-Dimensional Anyon-Hubbard Model}	

In synthetic systems, one can sometimes generate particles
which are neither bosons nor fermions, but behave as anyons with annihilaton
operator $a_{n}$ that have a statistical exchange phase $\theta$
which differs from both 0 and $\pi$, such that they satisfy a generalized
commutation relation
\begin{equation}
\left[a_{n},a_{m}^{\dagger}\right]_{\theta_{A}}\equiv a_{n}a_{m}^{\dagger}-e^{-i\theta_{A}\mathrm{sgn}(n-m)}a_{m}^{\dagger}a_{n}=\delta_{nm}
\label{eqn:anyonCommutator}
\end{equation}
where $\mathrm{sgn}(x)$ is the sign function,
$$
\mathrm{sgn}(n-m)=
\left\{\begin{array}{lr}
1,\, & n>m\\
0,\, & n=m\\
-1,\, & n<m
\end{array}\right\}.
$$
Note that these anyons acquire a phase $\theta_A$ when two particles
on different sites exchange places, but two particles on the same
site satisfy the normal bosonic commutation relation even for $\theta_{A}\neq0$.
A variety of recent proposals have suggested the possibility of generating effectively anyonic gases in one-dimensional ultracold atomic gases \cite{Keilmann2011,Greschner2015,Cardarelli2016,Strater2016}. 
A paradigmatic model describing such particles is the Anyon-Hubbard
model \cite{Keilmann2011,Greschner2015,Cardarelli2016,Strater2016}, in which these quasiparticles
with exotic exchange phases are governed by a typical Hubbard Hamiltonian,

\begin{equation}
\mathcal{H}_{A}=-J_{A}\sum_{n}\left(a_{n}^{\dagger}a_{n+1}+a_{n+1}^{\dagger}a_{n}\right)+\frac{1}{2}\sum_{n}U_{A}a_{n}^{\dagger}a_{n}^{\dagger}a_{n}a_{n}.
\label{eqn:Hanyon}
\end{equation}

The Hamiltonian for our system derived in the previous section can
be readily mapped onto the one-dimensional Anyon-Hubbard model. The
bosonic operators transform into anyon operators with the relationship,

\begin{equation}
a_{n}=\exp\left(i\frac{\theta_{A}}{2}N_{n}+i\theta_{A}\sum_{m=n+1}^{\infty}N_{m}\right)b_{n}.
\label{eqn:anyonOperators}
\end{equation}
With this transformation, simple algebra can verify that the effective
Hamiltonian maps onto the Anyon-Hubbard Hamiltonian
Eq.~(\ref{eqn:Hanyon}) and that the anyon operators Eq.~(\ref{eqn:anyonOperators})
satisfy the desired commutation relation Eq.~(\ref{eqn:anyonCommutator}). 

The parameters of the resulting Anyon-Hubbard model after the transformation are

\begin{eqnarray}
\theta_{A}\approx & -2\frac{J_{int}\cos(\Theta)}{J},\\ \label{eqn:thetaA}
J_{A}\approx & J,\\
U_{A}= & U_{0}, 
\end{eqnarray}
where the first two relationships rely on the assumption that the
statistical angle $\theta_{A}$ is small (specifically, $\tan(\theta_{A}/2)\approx\theta_{A}/2$; c.f. Eq.~(\ref{eqn:PeierlsPhase})).
Note that this limitation is not too stringent, especially in light of
the factor of two in Eq.~(\ref{eqn:thetaA}) (which arises because particles acquire the phase
$\theta_{A}/2$ when tunneling onto and again when tunneling away
from other particles). For example, one can set parameters to achieve
an anyon phase of $\theta_{A1}=\pi/4$ for atoms tunneling across
one neighbor at a time, and the effective phase per exchange $\theta_{A2}$
for tunneling across two neighbors simultaneously (i.e. tunneling
into, and subsequently out of, a site which was already doubly occupied)
would have an error of only $\theta_{A2}-2\theta_{A1}=2\arctan\left(2\tan\left(\theta_{A1}/2\right)\right)-2\theta_{A1}\approx-\pi/17$.
Further improvement in this respect can be achieved by taking advantage
of other interaction terms which add a negative real component to
the density-dependent tunneling, but the details of these corrections
are beyond the scope of this work. Alternatively, one can exactly
achieve arbitrary statistical angles (up to $\pi$) as long as it
is possible to neglect events in which particles tunnel onto sites
which were already occupied by two (or more) other particles; for example, this treatment would be valid in the limit of low density. 

The correspondence between the occupation-dependent Peierls phase and anyonic statistics is only valid in one dimension. When the system has multiple dimensions, particles can exchange locations without ever residing in the same site. In the one-dimensional chain, particles are forced to tunnel on top of each other in the process of exchanging locations, and thus the occupation-dependent tunneling phase is equivalent to a statistical exchange phase as shown.

Note that our method for simulating the Anyon-Hubbard model takes
an approach distinct from previous proposals; for example, while the
proposals in Ref.~\cite{Greschner2015,Strater2016} relied on a shaking protocol which induces photon-assisted tunneling
to modify the dynamics within the s-band while maintaining static interaction
strength, our proposal relies on creating a small admixture of the
p-band into the s-band and synchronizing the shaking lattice
with interaction strength modulation using a Feshbach resonance.

\end{document}